\documentclass[prl,twocolumn,preprintnumbers]{revtex4-1}%
\usepackage{epsf,epsfig}
\usepackage{amssymb,amsmath,amsfonts}

\def\KK{{\cal K}}

\def\NN{{\cal N}}
\def\OO{{\cal O}}

\def\qq{{\cal Q}}
\def\tts{{$tt^*$ }}

\def\Tr{{\rm {Tr}}}

\newcommand{\be}{\begin{equation}}
\newcommand{\ee}{\end{equation}}
\newcommand{\beq}{\begin{equation}}
\newcommand{\eeq}{\end{equation}}
\newcommand{\ben}{\begin{displaymath}}
\newcommand{\een}{\end{displaymath}}
\newcommand{\beqa}{\begin{eqnarray}}
\newcommand{\eeqa}{\end{eqnarray}}
\newcommand{\bea}{\begin{eqnarray}}
\newcommand{\eea}{\end{eqnarray}}
\newcommand{\bean}{\begin{eqnarray*}}
\newcommand{\eean}{\end{eqnarray*}}
\newcommand{\ba}{\begin{array}}
\newcommand{\ea}{\end{array}}
\newcommand{\bi}{\begin{itemize}}
\newcommand{\ei}{\end{itemize}}

{}{}
{}{}
\def\vereq#1#2{\lower3pt\vbox{\baselineskip1.5pt \lineskip1.5pt
\ialign{$\m@th#1\hfill##\hfil$\crcr#2\crcr\sim\crcr}}}
\makeatother

\begin{document}

\title{Exact correlation functions in $SU(2)$ ${\cal N}=2$ superconformal QCD}

\preprint{CCTP-2014-18}
\preprint{CCQCN-2014-42}
\preprint{CERN-PH-TH-2014-177}

\author{Marco Baggio}
\email{baggiom@ethz.ch}
\affiliation{ Institut fur Theoretische Physik, ETH Zurich, CH-8093 Zurich, Switzerland }

\author{Vasilis Niarchos}
\email{niarchos@physics.uoc.gr}
\affiliation{ Crete Center for Theoretical Physics
 and Crete Center for Quantum Complexity and Nanotechnology,
 Department of Physics, University of Crete, 71303, Greece}

\author{Kyriakos Papadodimas}
\email{ kyriakos.papadodimas@cern.ch \vspace{0.4cm}} 
\affiliation{Theory Group, Physics Department, CERN, CH-1211 Geneva 23, Switzerland, \\
on leave from the Centre for Theoretical Physics, University of Groningen,   The Netherlands}

\begin{abstract}
\noindent
We report an exact solution of 2- and 3-point functions of chiral primary fields in 
$SU(2)$ $\NN=2$ super-Yang-Mills theory coupled to four hypermultiplets. It is shown that 
these correlation functions are non-trivial functions of the gauge coupling, obeying differential equations
which take the form of the semi-infinite Toda chain. We solve these equations recursively in terms of the 
Zamolodchikov metric that can be determined exactly from supersymmetric localization
on the four-sphere. Our results are verified independently in perturbation theory with a Feynman
diagram computation up to 2-loops. This is a short version of a companion paper that contains detailed 
technical remarks, additional material and aspects of an extension to $SU(N)$ gauge group.
\end{abstract}
\maketitle

\section{Introduction}

Quantum field theories often possess exactly marginal deformations along which the data of the 
theory (spectrum, correlation functions, etc.) may change continuously. A characteristic well-studied
example in four dimensions is $\NN=4$ super-Yang-Mills (SYM) theory. In this case an exactly marginal
deformation interpolates between weak coupling (where the theory can be analyzed with standard 
perturbation theory) and strong coupling (where standard perturbative methods are inadequate).
It is of great interest to develop non-perturbative techniques that allow us to describe (analytically) 
properties of the theory at any value of the marginal couplings. 

Supersymmetric theories are an opportune context for the development of such techniques. 
They often possess special sectors that exhibit dynamics with non-trivial, but exactly computable, 
coupling constant dependence. An exact solution in these sectors can provide useful intuition, or
a solid starting point, towards an analysis of the more general properties of the theory.

In this note we will concentrate on a specific example of a four-dimensional conformal field theory with $\NN=2$
supersymmetry: $\NN=2$ SYM theory with gauge group $SU(2)$ coupled to 4 hypermultiplets in the fundamental
representation (in short, $SU(2)$ $\NN=2$ superconformal QCD, or simply SCQCD). By definition, this theory is 
invariant under 8 real supercharges. The special sector of interest comprises of (scalar) superconformal chiral
primary fields $\phi_I$ (to be specified explicitly in a moment) annihilated by the four supercharges of right chirality. 
The conjugate fields annihilated by the supercharges of left chirality will be denoted as $\overline{\phi}_{ I}$. 
$\NN=2$ superconformal field theories (SCFTs) are also invariant under the global $SU(2)_R\times U(1)_R$ 
R-symmetry. The chiral primaries $\phi_{I}$ are singlets of the $SU(2)_R$, but have non-zero $U(1)_R$ 
charge $R$ \cite{Dolan:2002zh}.
Their scaling dimension $\Delta$ obeys the relation $\Delta=\frac{R}{2}$. (For anti-chiral primaries 
$\Delta=-\frac{R}{2}$).

It is well known that the operator product expansion (OPE) of chiral primary fields is non-singular
\beq
\label{introaa}
\phi_I(x) \, \phi_J(0) = C_{IJ}^K \, \phi_K(0) +\ldots
~.
\eeq
It forms a ring structure known as the chiral ring \cite{Lerche:1989uy}. Two important sets of data in the chiral ring
are the 2-point functions
\beq
\label{introab}
\left< \phi_I(x) \, \overline{\phi}_{J}(0) \right> = \frac{g_{I\overline{J}}}{|x|^{2\Delta}}
\eeq
and the 3-point functions
\beq
\label{introac}
\left< \phi_I(x)  \phi_J(y)  \overline{\phi}_{ K}(z)\right>
= \frac{C_{IJ\overline{K}}}{|x-y|^{\Delta_{IJ, K}} \, 
|x-z|^{\Delta_{I K,J}} \,
|y-z|^{\Delta_{J K,I}}
},
\eeq
where
$\Delta_{IJ,K}=\Delta_I +\Delta_J -\Delta_K$.
There is an obvious relation between the OPE and 2- and 3-point function coefficients
$C_{IJ \overline{K}}= C_{IJ}^L\, g_{L\overline{K}}$.

In our example there is a single exactly marginal deformation labelled by a complex parameter 
$\tau$ (the complexified gauge coupling constant). 
The 2- and 3-point function coefficients $g_{I\overline{J}}$, $C_{IJ\overline{K}}$ are non-trivial functions
of $\tau$, receiving corrections at all orders in perturbation theory as well as from instanton effects.
(The scaling dimensions $\Delta_I$ are fixed by the non-renormalized $U(1)_R$ charge $R_I$
as described above). We will present exact formulae for these data combining methods of 
supersymmetric localization (in particular, \cite{Pestun:2007rz,Gerchkovitz:2014gta}) with certain exact relations 
between chiral ring correlation functions \cite{Papadodimas:2009eu}
that are four-dimensional analogs of the \tts equations in two dimensions \cite{Cecotti:1991me,Cecotti:1991vb}.
We have verified the resulting expressions with an independent computation in perturbation theory
up to 2-loops \cite{companion}.
 
We point out that analogous correlation functions in $\NN=4$ SYM theory are non-renormalized
\cite{Lee:1998bxa,D'Hoker:1998tz,D'Hoker:1999ea,Intriligator:1998ig,Intriligator:1999ff,Eden:1999gh,Petkou:1999fv,Howe:1999hz,Heslop:2001gp,Basu:2004nt,Baggio:2012rr} and are therefore trivial functions of the gauge coupling that 
can be determined at tree-level. $\NN=2$ dynamics is clearly more interesting and the results in this paper indicate 
that there is a considerable amount of new data that are tractable analytically compared to previous knowledge. 
The techniques presented here are useful in $\NN=2$ theories with exactly marginal deformations 
beyond the specific example analyzed in this note. A detailed explanation of the general properties of 
these techniques and extensions to more general examples are discussed in a companion paper \cite{companion}.

\section{$SU(2)$ ${\cal N}=2$ SCQCD}

The main example of this note is $\NN=2$ SYM theory with gauge group $SU(2)$ coupled to 4 hypermultiplets
(at the origin of the Coulomb branch).
This is a gauge theory whose field content includes: $(a)$ the $\NN=2$ vector multiplet fields, namely 
the gauge boson $A_\mu$, a complex scalar field $\varphi$ and four Weyl fermions (all in the adjoint 
representation); $(b)$ the 4 $\NN=2$ hypermultiplets that comprise of 4 complex bosons and 8 Weyl fermions
(all in the fundamental representation). The global symmetry group is $U(4)\times SU(2)_R \times U(1)_R$.
$U(4)$ is a flavor symmetry rotating the hypermultiplets. The standard Yang-Mills Lagrangian of this theory is 
summarized, for example, in appendix B of \cite{companion} whose conventions we are also following here.

The single exactly marginal coupling of this theory is the complexified Yang-Mills coupling 
$\tau=\frac{\theta}{2\pi} + \frac{4\pi i}{g_{YM}^2}$, where $\theta$ is the $\theta$-angle and $g_{YM}$
is the Yang-Mills coupling. We will work in conventions where the infinitesimal exactly marginal
deformation of the action takes the form
\beq
\label{sqcdaa}
S \to S +\frac{\delta \tau}{4\pi^2} \int d^4 x \, \OO_\tau(x) 
+\frac{\delta \bar \tau}{4\pi^2} \int d^4 x\, \overline{\OO}_{\tau}(x)
\eeq
where the $\Delta=4$ operators $\OO_\tau, \overline{\OO}_{\tau}$ are descendants of $\Delta=2$
(anti)chiral primary fields 
\beq
\label{sqcdab}
\OO_\tau = \qq^4\cdot \phi_2~, ~~ \overline{\OO}_{\tau} = \overline{\qq}^4 \cdot \overline{\phi}_2
~.
\eeq
The notation $\qq^4 \cdot \phi_2$ is shorthand notation for the nested (anti)-commutator of four supercharges
of left chirality. The Lorentz and $SU(2)_R$ indices of the supercharges are combined to give a Lorentz and
$SU(2)_R$ singlet. $\phi_2$ is the lowest dimension $\NN=2$ chiral primary field
\beq
\label{sqcdac}
\phi_2 = \frac{\pi}{8} \Tr [ \varphi^2 ]
~.
\eeq
The overall normalization in \eqref{sqcdab} is fixed so that 
$\left< \OO_\tau(x) \overline{\OO}_{\tau}(0) \right>= 
\nabla_x^2 \nabla_x^2 \left< \phi_2 (x) \overline{\phi}_2 (0) \right>$.

The chiral ring of the $SU(2)$ theory can be freely generated by the chiral primary field $\phi_2$ by 
repeated multiplication. The explicit checks reported below verify the consistency of this picture.  
We will normalize the generic chiral primary $\phi_{2n} \propto \left( \Tr [\varphi^2] \right)^n$ by requiring the OPE
\beq
\label{sqcdad}
\phi_2 (x) \, \phi_{2n}(0) = \phi_{2n+2}(0) +\ldots
~.
\eeq
This choice fixes all the non-vanishing OPE coefficients 
\beq
\label{sqcdae}
C_{2n~2m}^{2(n+m)} = 1
\eeq
and the normalization of all the higher order chiral primaries $\phi_{2n}$ $(n>1)$ which are multi-trace.

To summarize, the (chiral ring) sector of interest in this paper comprises of a sequence of fields $\phi_{2n}$
with scaling dimensions $\Delta_{2n} = 2n$.

We will denote the 2-point functions of these fields as
\beq
\label{sqcdaf}
\left< \phi_{2n}(x) \, \overline{\phi}_{2n}(0) \right> = \frac{g_{2n}(\tau,\bar\tau)}{|x|^{4n}}
~.
\eeq
The 2-point function coefficients $g_{2n}$ (as well as the corresponding 3-point function coefficients
$C_{2m\, 2n\, \overline{2m+2n}}$) are non-trivial functions of the complexified coupling $\tau$ that
we will determine exactly. 

Notice that $g_2$ is directly related to the coefficient $G_2$ of the 2-point function 
$\left< \OO_\tau(x) \overline{\OO}_{\tau}(0) \right>$. $G_2$ is the so-called Zamolodchikov
metric on the space of exactly marginal couplings. For $\NN=2$ theories this space is known to be
a complex K\"ahler manifold. Hence, (specializing to the case at hand) there is a scalar function $\KK$,
the K\"ahler potential, such that 
\beq
\label{sqcdag}
G_2 = \partial_\tau \partial_{\bar \tau} \KK = 192\, g_2
~.
\eeq

\section{Exact correlation functions}

Ref.\ \cite{Papadodimas:2009eu} formulated a set of exact relations between the OPE and 2-point function 
coefficients for general four-dimensional $\NN=2$ theories with exactly marginal directions. These relations,
which take the form of systems of differential equations on the marginal couplings, are direct analogs of the \tts 
equations in two-dimensional $\NN=(2,2)$ superconformal theories derived in \cite{Cecotti:1991me,Cecotti:1991vb} 
with the method of the topological-antitopological fusion. Ref.\ \cite{Papadodimas:2009eu} derived such relations in 
four dimensions with the judicious use of superconformal Ward identities.

Applying the general \tts equations of \cite{Papadodimas:2009eu} in the case of interest here 
in the so-called holomorphic gauge and the related above-mentioned normalization conventions (see 
\cite{companion} for an exposition of all the pertinent details) we arrive at the following relations for the 2-point 
function coefficients $g_{2n}$ \eqref{sqcdaf}
\beq
\label{exactaa}
\partial_\tau \partial_{\bar \tau} g_{2n} = \frac{g_{2n+2}}{g_{2n}} - \frac{g_{2n}}{g_{2n-2}} - g_2
\eeq 
where $n=1,2,\ldots$ and $g_0=1$ by definition. By unitarity all $g_{2n}>0$ and this infinite sequence of
differential equations can be recast as the more familiar semi-infinite Toda chain
\beq
\label{exactab}
\partial_\tau \partial_{\bar \tau} q_n = e^{q_{n+1}-q_n} - e^{q_n - q_{n-1}}~, ~~ n=2,\ldots
\eeq
by setting $g_{2n} = \exp\left( q_n - \log \left(\frac{\KK}{192} \right) \right)$. $\KK$ is the K\"ahler potential
in \eqref{sqcdag} and the factor of 192 follows from the normalization conventions of the previous section.

It is interesting to ask what is the general solution of the system \eqref{exactaa} subject to positivity over
the entire space of marginal couplings and whether positivity and some other `boundary conditions' from 
perturbation theory at weak coupling can fix the solution uniquely. We will not try to answer this question here. 
Instead, we will use the system of equations \eqref{exactaa} recursively, writing 
\beq
\label{exactac}
g_{2n+2} = g_{2n} \partial_\tau \partial_{\bar \tau} \log g_{2n} + \frac{g_{2n}^2}{g_{2n-2}}+g_2 g_{2n}
~, ~~ n=1,2,\ldots
~,
\eeq  
to determine all the higher 2-point functions $g_{2n}$ $(n>1)$ from the lowest one $g_2$.

\vspace{0.3cm}
\noindent
{\bf Exact 2-point functions.}
Recent work \cite{Gerchkovitz:2014gta} has determined the exact quantum K\"ahler potential of $\NN=2$ SCFTs in 
terms of the partition function $Z_{S^4}$ of the theory on the four-sphere $S^4$. The precise relation is
\beq
\label{exactba}
\KK = 192 \log Z_{S^4}
~.
\eeq
Notice that the marginal operators $\OO_\tau$ are normalized differently in \cite{Gerchkovitz:2014gta}, $i.e.$ 
$\OO_{here} = 4 \, \OO_{there}$. This explains the factor $192 = 12 \times 4 \times 4$ as opposed to 
12 in \cite{Gerchkovitz:2014gta}. Combining with \eqref{sqcdag} we obtain
\beq
\label{exactbb}
g_2 = \partial_\tau \partial_{\bar \tau} Z_{S^4}
~.
\eeq

For the $SU(2)$ SCQCD theory there is a well-studied integral expression for the sphere partition function
$Z_{S^4}$ that has been determined using supersymmetric localization \cite{Pestun:2007rz}
\begin{align}
\label{exactbc}
Z_{S^4}(\tau,\bar \tau)=
&\int_{-\infty}^\infty da~e^{-4\pi {\rm Im}(\tau) \, a^2} (2a)^2 
\nonumber\\
&\frac{H(2ia)H(-2ia)}{(H(ia)H(-ia))^4}
\left| Z_{\rm inst}(a,\tau) \right |^2
~.
\end{align}
$H(z)=G(1+z)G(1-z)$ in terms of the Barnes $G$-function \cite{Barnes}, and $Z_{\rm inst}$ is the Nekrasov
partition function \cite{Nekrasov:2002qd} that incorporates the contribution from all the instanton sectors. For further
details we refer the reader to \cite{Pestun:2007rz}. 

Combining the expressions \eqref{exactac}, \eqref{exactbb} and \eqref{exactbc} we are able to determine
recursively any of the 2-point function coefficients $g_{2n}$ in terms of higher derivatives of the $S^4$
partition function.

\vspace{0.3cm}
\noindent
{\bf Exact 3-point functions.} The non-vanishing 3-point function coefficients $C_{2m\, 2n\, \overline{2(m+n)}}$
follow immediately from the general relation $C_{IJ\overline{K}}= C_{IJ}^L\, g_{L\overline{K}}$, equation \eqref{sqcdae}, and
the above solution of the 2-point function coefficients
\beq
\label{exactca}
C_{2m\, 2n\, \overline{2(m+n)}} = C_{2m\, 2n}^{2(m+n)} \, g_{2(m+n)} = g_{2(m+n)}
~.
\eeq

Notice that, although the normalization conventions of the previous sections are very convenient for the 
above computations, in conformal field theory it is common to work instead with orthonormal primary operators 
$\hat \phi_{2n}$ for which 
$\left< \hat\phi_{2n}(x) \overline{\hat \phi}_{2\bar n}(0) \right> = \frac{\delta_{n,\bar n}}{|x|^{2\Delta}}$.
In these alternative conventions, the 2-point function coefficients are trivial but the OPE coefficients are
non-trivial and 
\beq
\label{exactcb}
\hat C_{2m\, 2n\, \overline{2m+2n}} = \sqrt{\frac{g_{2m+2n}}{g_{2m}\, g_{2n}}}
~.
\eeq

\vspace{0.3cm}
\noindent
{\bf More general extremal correlators.} With a conformal transformation of the form
${x'}^\mu = \frac{x^\mu-y^\mu}{|x-y|^2}$ it is possible to recast the general `extremal correlator'
\beq
\label{exactda}
\left< \phi_{2m_1}(x_1)\ldots \phi_{2m_n}(x_n)\overline{\phi}_{2\bar m}(y)\right>
\eeq
with $\bar m = \sum_{\ell=1}^n m_n$ as
\beq
\label{exactdb}
\frac{\left< \phi_{2m_1}({x'}_1)\ldots \phi_{2m_n}({x'}_n)\overline{\phi}_{2\bar m}(\infty)\right>}
{|x_1-y|^{4m_1} \cdots |x_n-y|^{4m_n}}
~.
\eeq
Using superconformal Ward identities one can prove that the correlation function on the numerator
of \eqref{exactdb} is independent of the positions $x_i$. Consequently, it can be evaluated in any particular
limit; in particular, we can make use of the above known OPEs and 2-point functions $g_{2n}$ to determine 
the exact $\tau$-dependence of such extremal correlators as well, as explained in more detail in \cite{companion}.

\section{Predictions for perturbation theory}

We can use the above results to make very specific predictions for the weak coupling, $g_{YM}\ll 1$, 
expansion of 2- and 3-point functions in the chiral ring. As an illustration, here we present explicit examples 
in the 0-instanton and 1-instanton sectors.

\vspace{0.1cm}
\noindent
{\bf 0-instanton sector.} Working with the perturbative (0-instanton) part of the $S^4$ partition function \eqref{exactbc}
\beq
\label{predictaa}
Z_{S^4}^{(0)} = \int_{-\infty}^\infty da~ e^{-4\pi {\rm Im}(\tau)\, a^2} (2a)^2 
\frac{H(2ia)H(-2ia)}{(H(ia)H(-ia))^4}
\eeq
our exact formulae provide, e.g. for the first three chiral primaries, the perturbative expansions
\begin{align}
\label{predictab}
g_2^{(0)} =& \frac38 \frac{1}{(\rm{Im} \tau)^2} - \frac{135 \,\zeta(3)}{32 \, \pi^2} \frac{1}{(\rm{Im} \tau)^4} + \frac{1575 \,\zeta(5)}{64 \, \pi^3} \frac{1}{(\rm{Im} \tau)^5} + \ldots, 
\nonumber \\
g_4^{(0)} =& \frac{15}{32} \frac{1}{(\rm{Im} \tau)^4} - \frac{945 \,\zeta(3)}{64 \, \pi^2} \frac{1}{(\rm{Im} \tau)^6} + \frac{7875 \,\zeta(5)}{64 \, \pi^3} \frac{1}{(\rm{Im} \tau)^7} + ..., 
\nonumber\\
g_6^{(0)} =& \frac{315}{256} \frac{1}{(\rm{Im} \tau)^6} - \frac{76545 \,\zeta(3)}{1024 \, \pi^2} \frac{1}{(\rm{Im} \tau)^8} 
\nonumber\\
&+ \frac{1677375 \,\zeta(5)}{2048 \, \pi^3} \frac{1}{(\rm{Im} \tau)^9} + \ldots
~.
\end{align}

The superscript $0$ denotes that this is the 0-instanton contribution. We wrote down contributions only up to 3-loops, but it is  easy to go to any desired order. We
have verified the validity of the predicted 
$g_{2n}^{(0)}$, for all values of the positive integer $n$, with an independent computation in perturbation theory
up to 2-loops \cite{companion}.  This provides an independent 2-loop perturbative check of the \tts equations 
\eqref{exactac}, but also a check of the recent proposal of Ref.\ \cite{Gerchkovitz:2014gta} that identifies the 
quantum K\"ahler potential of $\NN=2$ theories with the $S^4$ partition function.

Equivalently, in the alternative basis with orthonormal 2-point functions formula \eqref{exactcb} provides very 
specific results for the non-trivial 3-point function coefficients $\hat C_{2m\, 2n\, \overline{2m+2n}}$.
As an illustration the first few coefficients are
\begin{align}
\label{predictac}
\hat C_{2\, 2\, 4}^{(0)} & = \sqrt{\frac{10}{3}}\left(1 - \frac{9\,\zeta(3)}{2\pi^2} \frac{1}{(\rm{Im} \tau)^2} 
+ \frac{525 \, \zeta(5)}{8 \pi^3} \frac{1}{(\rm{Im} \tau)^3} + \ldots \right),
\nonumber\\
	\hat C_{2\, 4\, 6}^{(0)} & = \sqrt{7}\left(1 - \frac{9\,\zeta(3)}{\pi^2} \frac{1}{(\rm{Im} \tau)^2} 
	+ \frac{675 \, \zeta(5)}{4 \pi^3} \frac{1}{(\rm{Im} \tau)^3} + \ldots \right),
\nonumber\\
	\hat C_{2\, 6\, 8}^{(0)} & = \sqrt{12}\left(1 - \frac{27\,\zeta(3)}{2\pi^2} \frac{1}{(\rm{Im} \tau)^2} 
	+ \frac{2475 \, \zeta(5)}{8 \pi^3} \frac{1}{(\rm{Im} \tau)^3} + ... \right).
\end{align}

\noindent
{\bf 1-instanton sector.} As an example, we consider the 1-instanton contribution to the $S^4$ partition function \eqref{exactbc}
\begin{align}
\label{predictba}
&Z_{S^4}^{(1)} =  \cos\theta \exp\left(-{8 \pi^2 \over g_{_{YM}}^2}\right)\left(-{3 \over 4 \pi({\rm Im}\tau)^{3/2}}\right)
\\
&\Big[1- {1\over 8 \pi {\rm Im}\tau} - {45 \zeta(3) \over 16 \pi^2 ({\rm Im}\tau)^2} +{105(\zeta(3) + 10 \zeta(5)) \over 128 \pi^3 ({\rm Im}\tau)^3} + \ldots\big]
\nonumber
\end{align}
and from this we obtain expansions of $g_{2n}$ in the 1-instanton sector
\begin{align}
\label{predictbb}
& g_2^{(1)}  = \cos\theta \,\exp\left(-{8 \pi^2 \over g_{_{YM}}^2}\right)\Big[ {3\over 8 ({\rm Im}\tau)^2} 
+ {3 \over 16 \pi ({\rm Im}\tau)^3}
\nonumber\\ 
&  -{135 \zeta(3)\over32 \pi^2 ({\rm Im}\tau)^4} +\ldots \Big],
\\
& g_4^{(1)}  = \cos\theta \exp\left(-{8 \pi^2 \over g_{_{YM}}^2}\right) \Big[{15 \over 16 ({\rm Im}\tau)^4}
+ {15 \over 32 \pi ({\rm Im}\tau)^5}
\nonumber\\
& -{945 \zeta(3) \over 32 \pi^2 ({\rm Im}\tau)^6} +\ldots\Big]
~.
\end{align}
If desired it is straightforward to extend these results to higher $n$, higher instanton number $\ell$ and higher order in  the perturbative expansion around any given instanton sector.
It would be interesting to confirm them with an independent perturbative 
computation in the general $\ell$-instanton sector. Moreover, it would be interesting to verify the expected
positivity of the resulting expressions at general $n$.

\section{Outlook}

We reported exact non-perturbative formulae for 2- and 3-point functions of chiral primary fields in 
the $SU(2)$ $\NN=2$ SCQCD theory. A detailed exposition of the employed technology, of the 
perturbative 2-loop check, as well as an extension to the $\NN=2$ SCQCD theory with more general 
$SU(N)$ gauge group can be found in the companion paper \cite{companion}. Currently, we do not have a full
solution of the \tts equations in the $SU(N)$ case, but we find preliminary signs of an underlying structure
that remains to be understood better.

The present results indicate the possibility that the dependence of the chiral ring structure
of four-dimensional $\NN=2$ theories on their marginal couplings is exactly computable despite being 
highly non-trivial. It would be interesting to extend the application of the 4d \tts equations 
\cite{Papadodimas:2009eu} to other known classes of $\NN=2$ theories and to determine general 
conditions (e.g. positivity constraints) that fix their solution uniquely. Such solutions are expected to have wider 
implications. For example, we have already seen that the explicit knowledge of 2- and 3-point functions implies also 
the exact form of general extremal correlation functions in the chiral ring. In a different direction one can envision 
using these results as input in a more general bootstrap program in $\NN=2$ SCFTs aiming to determine larger 
classes of correlation functions, spectral data etc. Clearly, more remains to be done.

\vspace{0.3cm}
\begin{acknowledgments}
We would like to thank M. Buican, J. Drummond, M. Kelm, W. Lerche, B. Pioline, M. Rosso, D. 
Tong, C. Vafa, C. Vergu, C. Vollenweider and A. Zhedanov for useful discussions. The work of M.B. is supported in 
part by a grant of the Swiss National Science Foundation. The work of V.N. was supported in part by European 
Union's 7th Framework Programme under grant agreements (FP7-REGPOT-2012-2013-1) no 316165, 
PIF-GA-2011-300984, the EU program `Thales' MIS 375734 and was also co-financed by the EU 
(European Social Fund, ESF) and Greek national funds through the Operational Program `Education and Lifelong 
Learning' of the National Strategic Reference Framework (NSRF) under `Funding of proposals that have received a 
positive evaluation in the 3rd and 4th Call of ERC Grant Schemes'. K.P. would like to thank the Royal 
Netherlands Academy of Sciences (KNAW).
\end{acknowledgments}


\end{document}